# Electrokinetic Energy Harvesting using Paper and Pencil

Sankha Shuvra Das[a], Shantimoy Kar[b], Tarique Anwar[a], Partha Saha[a] and Suman Chakraborty[ab]*

[a]Department of Mechanical Engineering, Indian Institute of Technology Kharagpur, Kharagpur 721302, India.
[b]Advanced Technology Development Centre, Indian Institute of Technology Kharagpur, Kharagpur 721302, India.
*email: suman@mech.iitkgp.ernet.in

We exploit the combinatorial advantage of electrokinetics and tortutosity of cellulose-based paper network on a laboratory grade filter paper for the development of a simple, inexpensive, yet extremely robust (shows constant performance till 12 days) 'paper-and-pencil'-based device for energy harvesting application. We successfully achieve to harvest maximum output power of ~ 640 pW in single channel, while the same is significantly improved (by ~ 100 times) with the use of multichannel microfluidic array (maximum up to 20 channels). We envisage that such ultra-low cost devices may turn out to be extremely useful in energizing analytical microdevices in resource limited settings, for instance for extreme point of care diagnostics applications.

## Introduction

Throughout the last decade, paper-based microfluidics caught significant attention for myriad of applications[1–3], ranging from disease diagnostics[4–6], water quality control[7], food quality monitoring[8] to heavy metal ion detection[9]. Recent studies show that paper-based microfluidic devices may also be used for probing fundamental phenomena like micro-mixing[10], electro-wetting[11], digital microfluidics[12] etc. In most of the scenarios, paper-based devices have been utilized for qualitative/semi-quantitative purposes and have been preferred because of their frugality, disposability, and easy manufacturing[13]. In recent years, paper-based devices have also been found suitable for energy storage applications e.g. flexible electronics, development of fuel cells etc[14–18].

Miniaturization of various bio-electronics and microfluidic-based lab-on-chip devices has essentially demanded an integrated power source for powering those micro-chips[19]. Towards this end, realizing an alternate source of green energy generation for these microfluidic chips is certainly one of the key concerns. To achieve this feat, many different mechanisms like solar cells[20], dye sensitized solar cells (DSSCs)[21], bio-mass conversion[22], microbial energy harvesting[23] etc. are extensively explored. Despite the inherent advantages of the aforesaid processes, all the underlying principles have certain limitations. In parallel, electrokinetic energy conversion mechanisms, mediated by the establishment of a streaming potential (i.e. the potential generated due to the continuous transport of electrolytes), has of late emerged to be as effective alternative[24]. Recent studies have successfully demonstrated the application of this paradigm, albeit in a sophistically controllable laboratory environment that cannot possibly be replicated for catering the functionalities of point of care devices in resource-limited settings[25–30]. In addition, the reported devices on electrokinetic energy conversion necessitate elaborate device fabrication, expensive consumables and trained personnel. Furthermore, these devices not only demand very intricate operational module, but also do not inherently integrate with low cost analytical platforms (such as paper strips); which eventually makes the entire paradigm expensive[31].

Comprehensive literature review shows that recent endeavors have been directed towards facilitating on-chip power generation. Meng *et al.* illustrated the concept of electrochemically induced $CO_2$ bubble driven liquid fuel cell which seems to be useful for continuous fluid flow[32]. However, the associated fabrication difficulties and poor performance of the electrode restricts its utilitarian scope for continuous power generation. Arun *et al.* have demonstrated the usage of 'paper-and-pencil' based fuel cells for sustained period of power generation[33]. In such a system, graphite electrodes trap atmospheric oxygen and thus act as an internal source of oxygen. However, the process uses formic acid as a fuel and sulfuric acid as an oxidant which confine the scope of the device. Furthermore, performance enhancement of such devices (like microfluidic battery, triboelectric nanogenerators) requires a coating of different materials; hence, the efficiency is often constrained by the thickness of the adsorbed material[34–38].

In this work, we explore electrokinetics on a simple 'paper-and-pencil' based platform (shown in Figure 1) as a greener alternative for on-chip energy harvesting. The primary advantages of such a platform are the self-propelling nature of the input flow through an exploitation of intrinsic capillary transport in paper pores (to this end, no syringe pump or equivalent actuation is necessary), and an explicit integrability with paper based diagnostic platforms for point of care applications. These features empower the device with a favourable functionality in extremely challenging and resource limited settings in an ultra-low cost paradigm. As a consequence, the intrinsic porous capillaries in the paper structure drive a surface tension driven flow that induces ionic convection necessary for the establishment of an electrical potential across the device, resulting in a favourable direct exploitation and conversion of surface energy into electrical power.

We use standard laboratory grade filter paper (whatmann grade 1). We achieve to successfully harvest power of ~ 640 pW in single channel; which is further improved by O (~) $10^2$ by connecting multiple channels (maximum up to 20 channels) through series connection. Our approach delineates a frugal, efficient and yet extremely robust platform for energy harvesting for sustained duration (up to 12 days) without requiring any sophisticated laboratory environment.

## Experimental details

**Chemicals**

Potassium chloride (Merck Life Science Private Ltd.) and Whatman cellulose filter paper (GE Healthcare, UK) of grade 1 are used for the experimentation. Electrolyte solutions are prepared by mixing KCl in Milli Q deionized water (18MΩ cm).

**Device fabrication**

Paper channels are fabricated using photolithography technique similar to the process demonstrated by Mandal *et al*[39] (schematically delineated in Figure S1; see ESI). Hydrophobic coating of the paper leads to the blockage (evident from Figure S2; see ESI) of the paper pores and therefore the fabricated hydrophobic barrier guides the fluid transportation in definite direction. To fabricate the graphite electrodes, reservoir pads are sketched using HB pencil (Figure S2 in ESI). Silver wire (Sigma Aldrich: ≥99.99% trace metal basis, 1.59 µΩ-cm at 20°C) of ~250 µm diameter is attached to the pencil sketched electrodes using conductive silver paste (Alfa Aesar) for the measurement of the potential. To measure the streaming potential, Keithley 2182A nanovoltmeter is connected in parallel with the electrodes. Following the analogy of electrokinetics (i.e. the migration of counter ions in downstream direction), higher end of the nanovoltmeter probe is connected to downstream electrode (outlet reservoir pad) and the lower end is connected to the upstream electrode (inlet reservoir pad). In order to acquire the data continuously from nanovoltmeter, we use an in-house lab-view code. Details of the experimental and data measurement setup are schematically delineated in Figure 2. We use 1 mM KCl solution as the electrolyte throughout the entire course of investigation (for details see Figure S3 in ESI). To eradicate the effect of evaporation, uniform experimental condition (relative humidity: 50%, temperature: 22-24$^0$C) is maintained throughout the course of study.

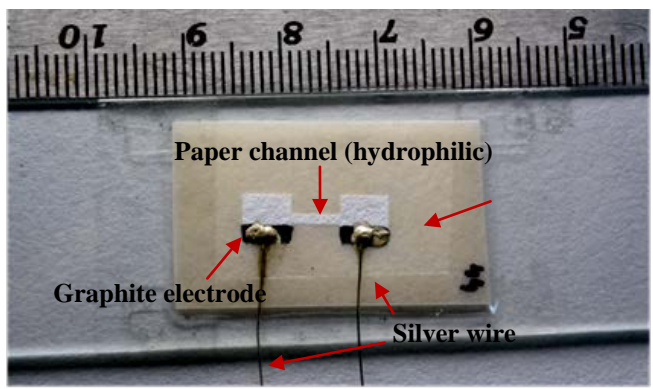

**Figure 1:** 'Paper-and-pencil'-based device used for streaming potential measurement. Device dimensions are 1 mm wide, 5 mm long with square inlet and outlet reservoir pads of 6mm × 6mm dimensions.

**Equivalent electrical circuit**

We exploit an analogy of equivalent electrical connection (shown in Figure 3) for continuous measurement of streaming potential and output power. Paper channel (i.e. the hydrophilic part of the device) consists of numbers of micropores which can be assumed as an array of microchannels. We connect a nanovoltmeter in parallel with the electrodes to measure the open circuit potential. Further, an external resistance ($R_L$) is connected in parallel to the circuit to measure the respective close circuit potential. Inherent presence of the capillary force drives the electrolyte solution to the downstream direction through the micropores with resistance $R_{micropore}$.

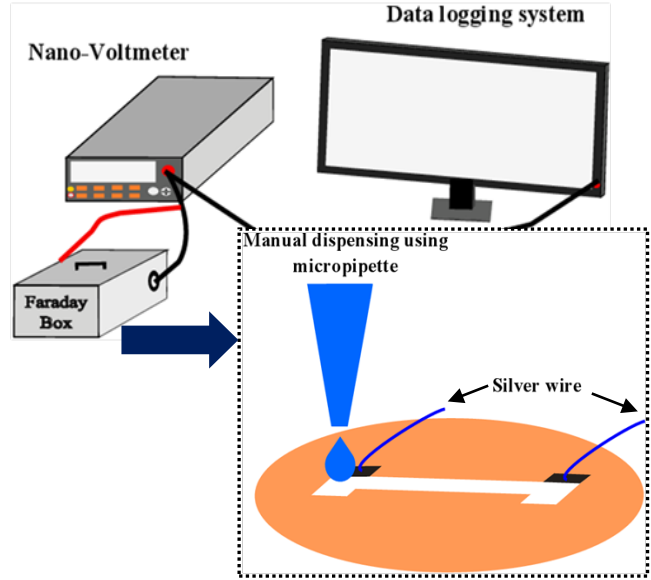

**Figure 2:** Schematic representation of the experimental setup for measuring streaming potential on 'paper-and-pencil' platform. Inset shows the inside view of Faraday box with manual dispensing of KCl solution on loading pad of the channel.

$I_{str}$ is the streaming current that is being generated due to the advection of ions which develops an electrical field due to ion transportation and is known as streaming potential. This, in turn, induces the conduction current, $I_c$ acting opposite to $I_{str}$. The open circuit streaming potential can be calculated as[40]:

$$\Delta V_{stro} = Z(I_{str} - I_c)R_{micropore} \qquad (1)$$

where Z is the number of micropores within the channel area. However, in case of close circuit condition it can be calculated as,

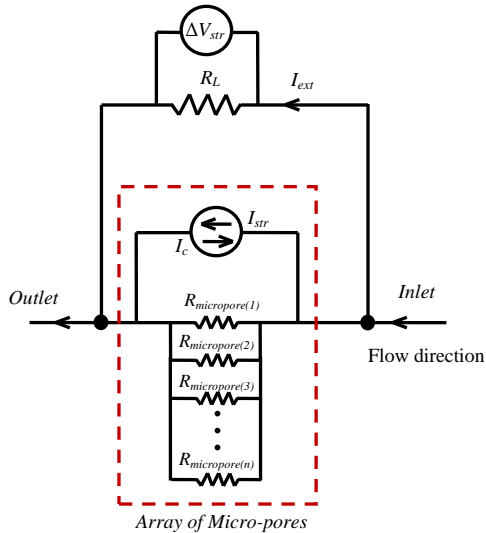

**Figure 3:** Schematic of equivalent electrical circuit used for streaming potential and output power measurement.

$$\Delta V_{strc} = I_{ext} R_L \tag{2}$$

$$\text{and } I_{ext} = Z(I_{str} - I_c) \tag{3}$$

Now, the output power ($P_{output}$) can be calculated using the following formula:

$$P_{output} = \frac{(\Delta V_{strc})^2}{R_L} \tag{4}.$$

## Results and discussions

Temporal variation of the induced open circuit potential is shown in Figure 4. Initial fluctuations can be ascribed to the fact of incomplete circuit (i.e. fluid is yet to reach to the outlet reservoir pad). Once the fluid makes contact with the outlet reservoir electrode, there is a sharp rise in the measured potential and thereafter it gradually decreases before it reaches a plateau regime.

Fundamental understanding about the microscopic structures of the paper and subsequent flow characteristics through such tortuous network is essential to explain the observed behavior of the streaming potential. Overall flow mechanism through paper-based microfluidic devices is fundamentally different from other conventional microfluidic platforms (glass, silicon, PDMS-based platforms). Due to the presence of inherent capillary action, fluid imbibes through the porous network of the paper matrix, not flowing on top of the surface. Paper is composed of enormous number of cellulose fabrics which are randomly distributed over the entire surface. Due to the presence of free carboxylic acid and hydroxyl groups, cellulose fibres are known to have free negative charges on its surface in contact with KCl solution, which is confirmed by the measured zeta potential (-8.76 ± 0.7813 mV)[41].

Therefore, due to the generation electrical double layer (EDL) on cellulose fabrics (effect of electrolyte's concentration is illustrated in Figure S3, please see ESI), there will be surplus of the counter ions in the downstream part of the channel (i.e. towards the direction of flow) which essentially leads to the generation of streaming potential across the two ends of the device (schematically shown in Figure 5). Interestingly, this occurs at virtually no expense, since the intrinsic porous capillaries in the paper structure essentially drive a surface tension driven flow that induces the advective transport of the ionic species for the establishment of the streaming potential, and no external pumping mechanism such as the syringe pump becomes necessary. Furthermore, the induced potential is certainly to be highest at the very initial phase i.e. when the circuit completes the connection for the first time and thereafter the potential drop is gradually decreasing. From Figure 4, it is clearly understood that at the very initial phase, the highest potential is measured which is continuously decreased with time and finally being stable for more than 2 hours; which indicates the fact that there is no more significant transportation of ions in macroscopic scale (i.e. in microscopic scale, ions are migrating in equivalent rate in almost all possible directions).

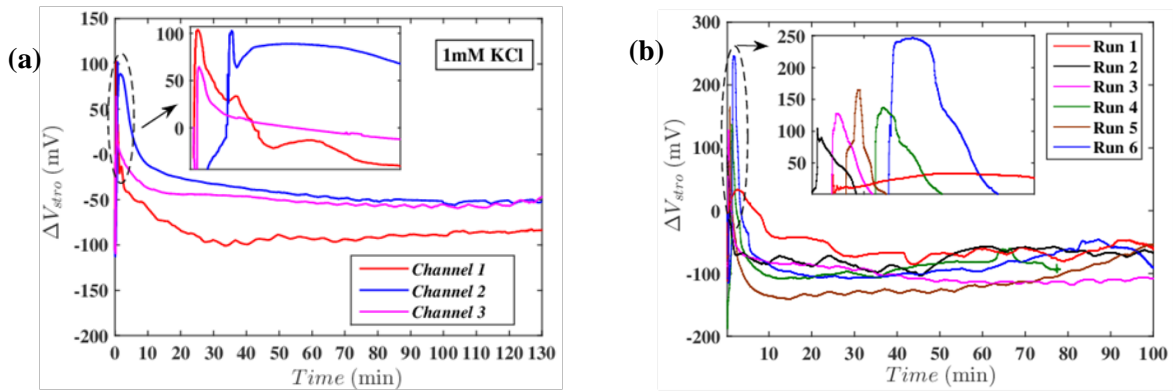

**Figure 4:** Temporal measurement of induced open circuit streaming potential for 1mM KCl solution in **(a)** three different channel, **(b)** same channel for six consecutive runs [insets show the maximum open circuit potential in each run].

From Figure 4a, it can be inferred that the reproducibility of the measured results differs in different channels which can be accounted from the random distribution of the cellulose fabrics on the paper matrix. In Figure 4b, experimental reproducibility on the same channel is presented. Initial priming with KCl solution certainly decreases (as the micropores have already developed EDL) the initial fluctuation, whilst similar trend is observed for sustained period of measurement.

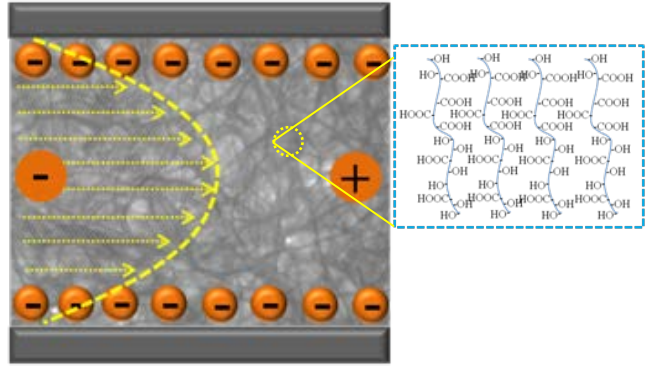

**Figure 5:** Schematic of the streaming potential generation on cellulose fibres (zoomed view: cellulose threads have –OH and –COOH functionality). Arrows indicate the direction of flow.

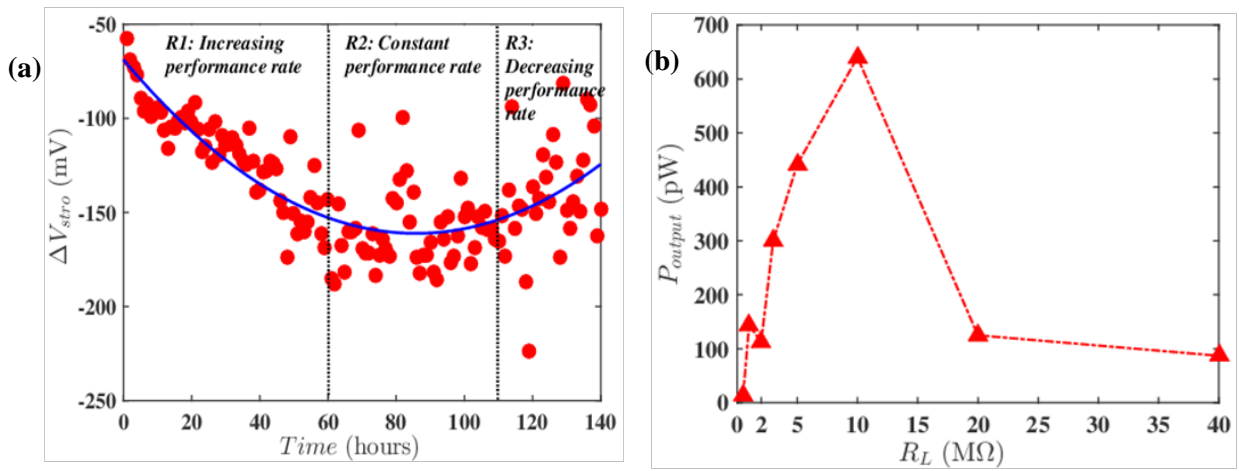

**Figure 6: (a)** Cyclic test to investigate the performance of the paper channel with time and **(b)** variation of output power with respect to different load resistances (connected parallel to the circuit).

Following the results illustrated in Figure 4, it is clearly evident that the random distribution of cellulose network imparts some degree of fluctuation in measurements (particularly for longer period of the experiments). In this context, one obvious concern arises what is the life span of the device? To address this particular concern, we pay attention to understand the life cycle of the device. We perform a cyclic test which consists of 12 hours continuous measurement followed by ~10-12 hours of drying of the channel prior to the next cycle. The KCl solution (~50μl) is dispensed on the loading pad at an interval of one hour. From Figure 6a, three distinct regimes are observed. Regime 1 (R1) indicates the constant increase of performance rate of the device till ~60 hours, whereas the R2 indicates almost constant performance rate till ~100-110 hours; in R3 delineates attenuation of performance rate after 110 hours. The highest $\Delta V_{stro}$ (measured open circuit potential) obtained as ~ -190mV at ~90 hours. So, it is important to note here that the device performance remains same even after ~140 hours of continuous operation; which is certainly an additional benefit for long term applications.

To measure the closed circuit potential, an external load ($R_L$) is connected in parallel to the circuit. The variation of calculated output power agaisnt different load resistance is delineated in Fig 6b. The maximum output power for single channel is measured to be ~ 640 pW for the external resistance of 10 MΩ.

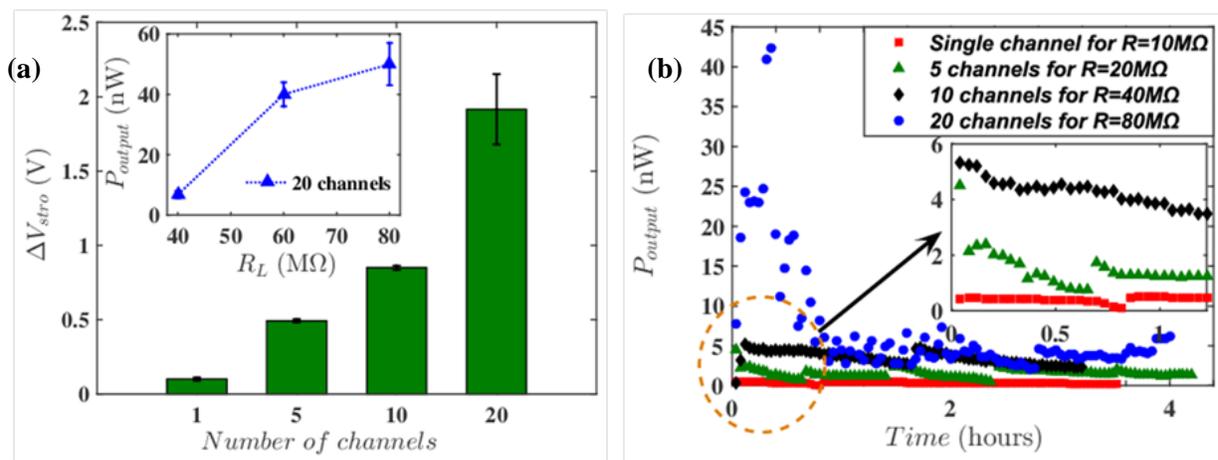

**Figure 7:** Variation of **(a)** open circuit potential for multichannel array shows the increase of open circuit potential as the number of channels in array is increasing from 1 to 5 to 10 to 20, (inset shows how the output power differs against different external load for 20 channels); and **(b)** temporal evolution of output power with respect to optimum external load resistance for multichannel array (inset indicates that at initial time points measured output power is higher while at later phase (~after 1 hour) it stabilizes to a constant potential value.

To further improve the output power, we connect the multiple channels (maximum up to 20 channels) through series connection (Figure S4 in ESI shows 5 channels; while the supplementary video shows a detailed measurement setup). Figure 7a shows that the maximum $\Delta V_{stro}$ observed is ~ 100mV for single channel, ~ 482mV for 5 channels, ~ 862mV for 10 channels and ~ 2.1V while 20 channels are in series connection. In order to understand the device performance, the output power for single channel as well as for different channel combinations against different external loads is measured (depicted in Figure S5, see ESI). From this device optimization study, it is seen that for single channel maximum $P_{Output}$ is seen against $10\,M\Omega$, whereas it is against $20\,M\Omega$, $40\,M\Omega$ and $80\,M\Omega$ for channel combination 5, 10, and 20 respectively. The temporal evolution of the output power for single channel and multiple channels (measured against the specific external load for which it shows maximum power) is seen to be almost constant as seen in Figure 7(b) for a period of 4 hours. Moreover, the device performance is seemed to be excellent till 12 days of measurements (see ESI, Figure S6); which indicates the robustness of the device.

## Conclusions

In summary, a simple and frugal 'paper-and-pencil' based microfluidic array (on normal laboratory grade filter paper) has been demonstrated to be an efficient energy harvesting platform, having an inherent integrability with a corresponding analytical component (such as point of care diagnostics) of the same platform in an ultra-low cost paradigm. This virtually necessitates no input power that makes it potentially effective in challenging environments for point of care applications as hallmarked by resource-limited settings. The device essentially demonstrates a conceptualization of electrokinetics (streaming potential) in a paper based microfluidic platform, aided by an intrinsic surface tension driven connective transport of the ionic species through the porous network of the paper matrix. The device with single channel can develop maximum output power of ~640pW against external load of 10MΩ, whilst the maximum open circuit potential is observed ~190mV. Multi-channel microfluidic arrays implemented on the same platform show substantial improvement in the measured output power (enhanced up to ~50nW by connecting 20 channels in series, for demonstration). Moreover, we perform comprehensive study to understand the optimized performance and robustness of the device. Cyclic test of the device revealed that $\Delta V_{stro}$ as well as the efficiency increases with the number of usages of the device and shows almost constant efficiency up to 12 days.

We envisage that these simple power generation platforms will find its application in the development of green battery, can further be integrated power source for MEMS based low powering devices, and for self-powering paper-based medical diagnostic platforms that hold the potential of revolutionizing the implementation of point of care diagnostics in resource limited settings.

## Acknowledgements

The authors would like to acknowledge Sponsored Research and Industrial Consultancy (SRIC) cell, IIT Kharagpur for the financial support to the 'Plant on-a-Chip' project provided through the SGDRI grant.

## Notes and references

# Electronic Supplementary Information (ESI)

# 'Electrokinetic Energy Harvesting using Paper and Pencil'

Sankha Shuvra Das[a], Shantimoy Kar[b], Tarique Anwar[a], Partha Saha[a] and Suman Chakraborty[ab]*

[1]Department of Mechanical Engineering, Indian Institute of Technology Kharagpur, Kharagpur 721302, India.
[2]Advanced Technology Development Centre, Indian Institute of Technology Kharagpur, Kharagpur 721302, India.
email: suman@mech.iitkgp.ernet.in

## 1. Fabrication details

Whatman filter paper (laboratory grade 1; mean pore diameter ~11µm and thickness ~100µm) are used for our study. Initially, the filter paper was soaked homogeneously in negative photoresist (SU-8 10; MicroChem) for 2-3 minutes. The excess photoresist was squeezed out from the paper to ensure uniform spreading throughout the paper. Thereafter, the photoresist soaked paper was prebaked at 130°C for ~10 minutes to evaporate the excess solvent. In the subsequent steps, the paper was cooled to room temperature and exposed under UV radiation ($\lambda$~365nm) at 100 mW/cm$^2$ for 20 seconds through a positive mask using mask aligner (OAI Hybralign 2000 series). UV-exposed paper is further post-baked at 130°C for ~10 minutes followed by soaking in acetone for ~2 minutes (for development) and rinsed in Isopropanol. Finally, the developed channel was dried at 75°-80°C for ~30 minutes.

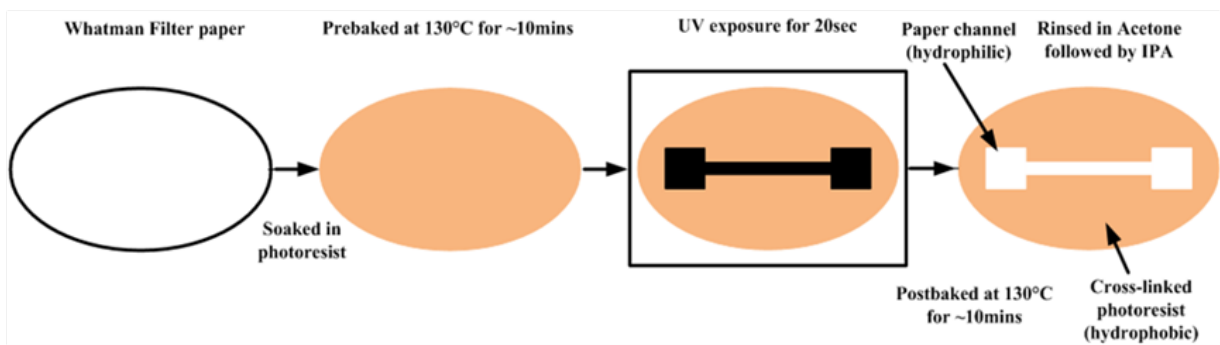

**Figure S1:** Schematic representation of fabrication of the paper microchannel.

## 2. Fabrication of pencil-sketched graphite electrodes:

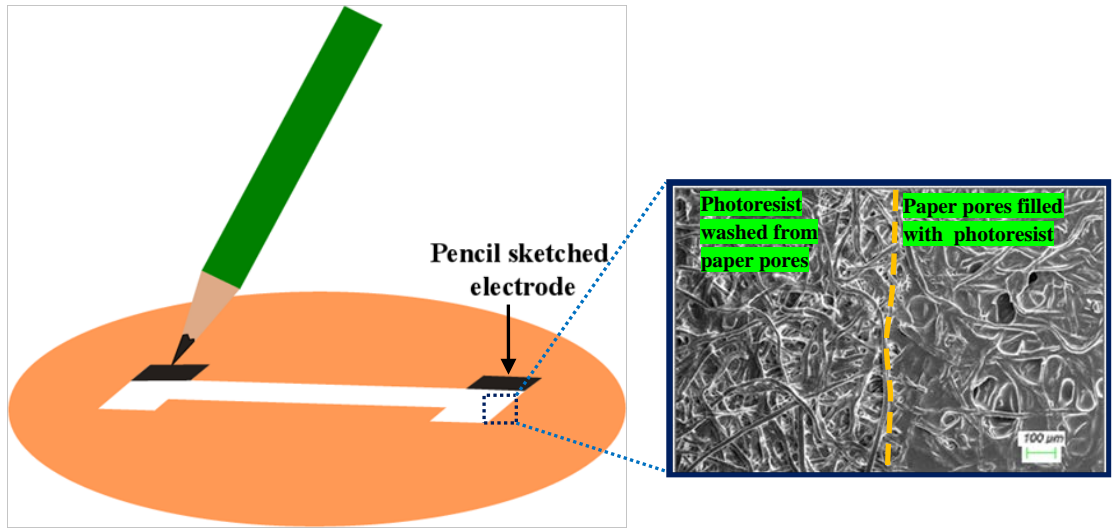

**Figure S2:** Schematic representation of electrode fabrication (pencil was sketched on front side of the paper surface) and scanning electron micrographs of the fabricated paper device. Half of the reservoir pads were kept free of graphite for the ease of fluid transportation.

## 3. Effect of electrolyte concentration

The thickness of electrical double layer (EDL) is known to vary with the concentration of electrolyte solution[1,2]. Therefore, we investigate the effect of electrolyte concentration to explore the effect of EDL on induced streaming potential on the 'paper-and-pencil'-based platforms.

The induced potential for 10mM and 100mM KCl solution is ~ -25mV and ~ -5mV respectively. Thus the induced streaming potential for 10mM and 100mM KCl solution is 4 times and 20 times less than as compared to 1mM KCl solution (~ -100mV) respectively. Hence, we chose 1mM KCl as the working electrolyte for our experimentations.

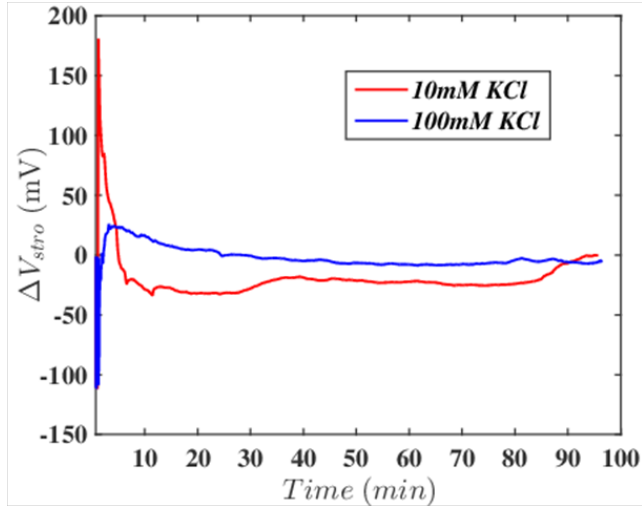

**Figure S3:** Temporal variation of induced open circuit streaming potential for 10mM and 100mM KCl solution.

## 4. Multichannel microfluidic array

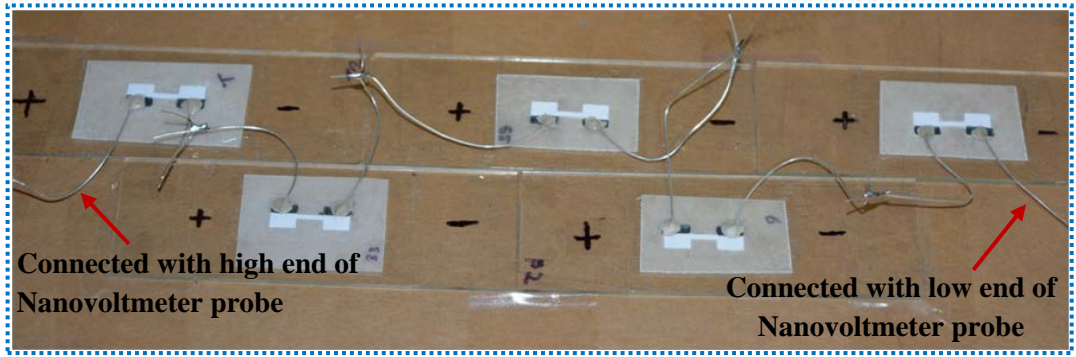

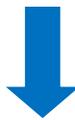

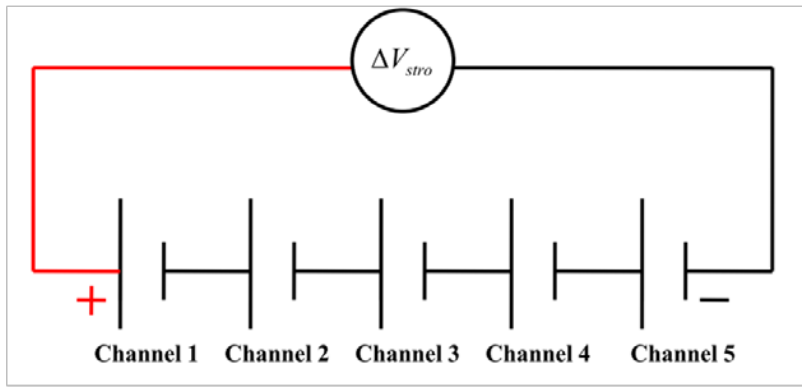

**Figure S4:** Series connection of multiple channels (for 5 channels).

## 5. Measurement of device performance against external loads

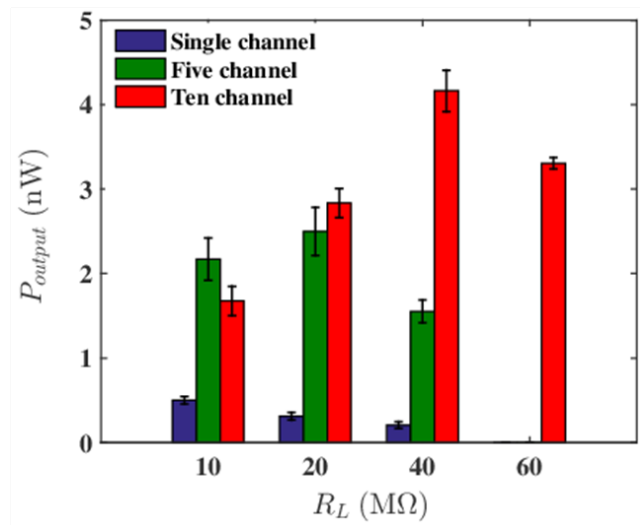

**Figure S5:** Variation of output power against different load resistance values for different channels.

## 6. Device performance in different days

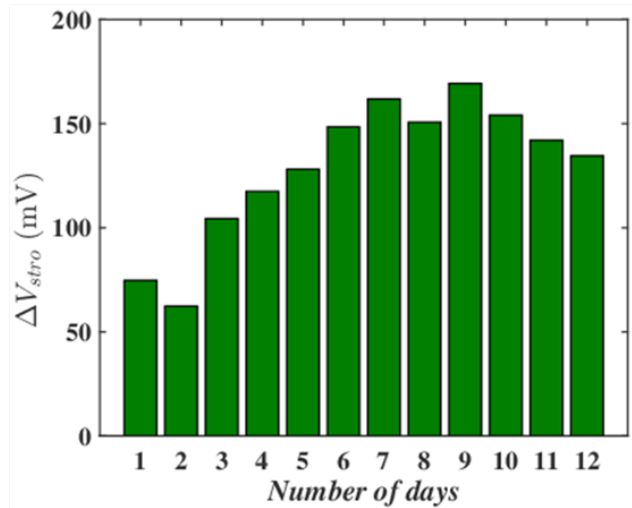

**Figure S6:** Variation of streaming potential for experiment conducted in different days (for same device with dispensing of electrolyte solution in each ~1 hours; for single channel).